\documentclass[
]{elsarticle}

\usepackage{amsmath}
\usepackage{amssymb}
\usepackage{amsfonts}
\usepackage{mdwmath}
\usepackage{mdwtab}

\usepackage{subcaption}

\usepackage{tikz}

\usepackage{algorithmicx}
\usepackage[noend]{algpseudocode}

\usepackage{url}
\usepackage{paralist}
\usepackage{microtype}
\usepackage{varwidth}

\title{%
  Enabling Interactive Mobile Simulations Through Distributed Reduced
  Models
}

\author[ipvs]{Christoph~Dibak}
\ead{dibak@ipvs.uni-stuttgart.de}

\author[ians]{Bernard~Haasdonk}
\ead{haasdonk@mathematik.uni-stuttgart.de}

\author[ians]{Andreas~Schmidt}
\ead{schmidta@mathematik.uni-stuttgart.de}

\author[ipvs]{Frank~D{\"u}rr}
\ead{duerr@ipvs.uni-stuttgart.de}

\author[ipvs]{Kurt~Rothermel}
\ead{rothermel@ipvs.uni-stuttgart.de}

\address[ipvs]{%
  Institute of Parallel and Distributed Systems\\
  University of Stuttgart,
  Stuttgart, Germany\\
  Email: \{dibak, duerr, rothermel\}@ipvs.uni-stuttgart.de
}

\address[ians]{%
  Institute of Applied Analysis and Numerical Simulation\\
  University of Stuttgart,
  Stuttgart, Germany\\
  Email: \{haasdonk, schmidta\}@mathematik.uni-stuttgart.de
}

\algdef{SE}[DOWHILE]{Do}{doWhile}{\algorithmicdo}[1]{\algorithmicwhile\ #1}

\makeatletter
\algrenewcommand\ALG@beginalgorithmic{\small}
\makeatother

\makeatletter
\def\ps@pprintTitle{%
  \let\@oddhead\@empty%
  \let\@evenhead\@empty%
}
\makeatother

\begin{document}

\begin{frontmatter}

  \begin{abstract}
    Currently, various hardware and software companies are developing
augmented reality devices, most prominently Microsoft with its Hololens.
Besides gaming, such devices can be used for serious pervasive
applications, like interactive mobile simulations to support engineers
in the field. Interactive simulations have high demands on resources,
which the mobile device alone is unable to satisfy. Therefore, we
propose a framework to support mobile simulations by distributing the
computation between the mobile device and a remote server based on the
reduced basis method. Evaluations show that we can speed-up the
numerical computation by over 131 times while using 73 times less
energy.

  \end{abstract}


\end{frontmatter}

  \begin{tikzpicture}[remember picture, overlay]
    \node[rectangle, draw, anchor=south, text centered]
      at ([yshift=4em] current page.south) {
      \begin{varwidth}{\textwidth}
        \begin{footnotesize}
            \textcopyright\,2018.  This preprint manuscript version is made available under the CC-BY-NC-ND 4.0 licence
            \url{https://creativecommons.org/licenses/by-nc-nd/4.0/}

            The formal version can be found at
            \url{https://doi.org/10.1016/j.pmcj.2018.02.002}
        \end{footnotesize}
      \end{varwidth}
    };
  \end{tikzpicture}

\section{Introduction}
\label{sec:introduction}

\noindent
Currently, several hardware and software companies announced or already
released augmented reality devices, such as Microsoft's Hololens, Sony's
SmartEyeglass, or Epson's Moverio Pro.  Using such devices allows the
user to see the real world augmented with additional information, such
as the state of the object the user is looking at or even holograms
displaying 3D scenes over the real world.

One interesting application of such devices is to display the result of
numerical simulations.  Having simulation results ubiquitously available
supports engineers or decision makers in the field~\cite{Dibak2014,
Dibak2015, Dibak2017b}.  As an example for such pervasive applications,
consider an engineer who has to find a solution for placing a hot
exhaust tube during deployment of a machine in a factory. To this end,
the engineer uses her augmented reality device, which directly shows the
heat of the surface of the tube as if the machine were operational.  She
can adjust the airflow surrounding the tube by changing parameters. The
application enables her to see the heat even in complex regions, e.g.,
in bends, and she can place the tube according to surrounding materials.
Other applications for mobile simulations are the visualization of
simulation results based on readings from nearby sensors for Internet of
Things applications or simulations on drones in order to exploit the
wind-field for energy-efficient flight routes~\cite{Ware2016}.

The main challenge for the computation of simulation results on mobile
devices is the computational complexity.  While it is feasible to
visualize even complex 3D data using dedicated GPUs on the mobile
device, calculating this data is still challenging due to slow
processors and limited energy resources of mobile devices.  Therefore,
we present an approach to support interactive simulations on
resource-constrained mobile devices by distributing computations between
the mobile device and a server infrastructure.  Moreover, the programmer
should be able to re-use existing simulation code.  Such code is
typically optimized for the server architecture, which might include
special hardware, such as additional GPUs, or vector processing
instruction sets, not available on the mobile device.  Besides the
computational aspect, the communication overhead is critical for an
efficient distributed execution on mobile devices and a remote server.
Therefore, we also need to minimize the amount of data communicated
between mobile device and~server.

To support numerical simulations on mobile devices, we present a
middleware in this paper that allows for the efficient distribution of
simulations between a mobile device and remote server infrastructure.
This middleware is based on the so-called Reduced Basis Method (RBM)
that allows for reducing the complexity of simulations by
pre-calculating an optimized basis reducing the dimensionality of the
simulation problem.  The basic idea of applying the RBM to distributed
mobile simulations is to calculate the reduced basis remotely on a
server and to transfer the basis to the mobile device.  Using the reduced
basis, the mobile device is able to perform simulations locally with
reduced computational overhead to calculate approximate solutions, where
the reduction of quality of approximate solutions is well-defined and
bounded. 

Using the RBM for efficient distributed simulations is not completely
novel and has already been proposed in~\cite{Huynh2010}. However,
approaches so far are rather static. They calculate and deploy a basis
once, which is then used for all subsequent simulations.  In contrast,
our approach allows for the dynamic adaptation of the basis during
runtime.

In detail, the contributions of this article are:\ %
\begin{inparaenum}[(1)]
  \item identification of a new class of serious augmented reality
    applications, namely interactive mobile simulations;
  \item presentation of a mobile simulation middleware that supports the
    programmer in implementing such applications;
  \item two approaches to enable mobile interactive simulations with and
    without a-priori known parameter restrictions;
  \item two approaches optimizing the bottleneck on mobile devices,
    which is reading data from internal storage;
  \item one approach to improve the pre-calculation step of the RBM for
    our mobile approach;
  \item real-world evaluations, including comparison of run-time and
    energy consumption of the methods, showing that our approaches are
    up to over 131 times faster and consume 73 times less energy
    compared to offloading everything to a server.
\end{inparaenum}

This work is an extension of our prior work~\cite{Dibak2017a}, in which
we presented two approaches to enable mobile interactive simulations and
one approach to reduce the amount of data to be read from internal
storage.  In this work, we extend our previous work by another approach
for reducing the read operations which is built on top of the previously
introduced approach.  Additionally, we present a novel approach for the
optimized pre-calculation on the server that further reduces the energy
cost during runtime on the mobile device.

The rest of the paper is structured as follows.  Section~\ref{sec:rbm}
provides background on the RBM followed by the system model in
Section~\ref{sec:model}.  Section~\ref{sec:basic} introduces a basic
approach, which will be modified to the adaptive approach in
Section~\ref{sec:adaptive}.  In order to reduce the number of snapshots
during the computation, we present the subspace approach and the reorder
approach in Section~\ref{sec:subspace} and Section~\ref{sec:reorder}.
Section~\ref{sec:reorder-basis-generation} introduces a new approach for
pre-processing on the server in order to improve the reorder approach.
Section~\ref{sec:evaluation} evaluates our approaches, before we discuss
related work in Section~\ref{sec:relatedwork} and conclude the paper.


\section{The Reduced Basis Method}
\label{sec:rbm}

\noindent
Our approach to enable interactive mobile simulations utilizes the
Reduced Basis Method (RBM) to solve complex numerical problems by
calculating parts of the simulation on a remote server infrastructure.
In order to better understand the problem to be solved and the solution,
we first give a brief introduction to the numerical problems to be
solved and the RBM in this section, before we explain the approach in
the following section.  A more in depth explanation of RBM can be found
in~\cite{Patera2007, Haasdonk2011b, Haasdonk2014}.

\subsection{The Full Numerical Problem}

\noindent
Simulations predict the behavior of a system based on a model. Commonly,
such models are described using differential equations.  Such equations
need to be discretized, which leads to large algebraic equations of the
form $A \cdot u = f$, where $A$ is a given matrix, $f$ is a given vector
called the right-hand side, and $u$ is the unknown solution.  We call
this equation the \emph{full problem}.

Simulation models contain parameters describing different properties of
the system that can be changed.  Such parameters can be used to interact
with the system, e.g., to insert sensor readings or user input.  To
express the dependency on parameters, we formulate the full problem as
\begin{align}
  \mathbf{A}(\mu) \cdot \mathbf{u}(\mu) = \mathbf{f}(\mu)
  \label{align:full-problem}
\end{align}
where $\mu$ is a vector including all parameters of the simulation.

\subsection{Parameter Separable Matrices}
\label{subsec:separable-matrices}

\noindent
The essential part of RBM is the parameter separability of the
matrix $A(\mu)$ and the right-hand side $f(\mu)$.  Parameter
separability of a vector or matrix $M$ is given if we know scalar
functions $\theta_i$ that map from the parameter space to real numbers
and matrices $M_i$ that have the same shape as $M$ such that
\begin{align}
  \mathbf{M}(\mu) = \sum_i \theta_i(\mu) \mathbf{M}_i. \label{alg:separability}
\end{align}
Such a representation can be derived from the model equation or using
Empirical Interpolation Methods~\cite{Barrault2004}.

\subsection{The Reduced Problem}
\label{ssec:reduced-problem}

\noindent
The RBM represents approximate solutions of the full numerical problem
as linear combination of \emph{snapshots}.  Snapshots are pre-computed
and linear independent solutions for typical parameters.  The snapshots
form the snapshot matrix $V$.  Therefore, the approximation to the real
solution $u(\mu)$ is $V u_V(\mu) \approx u(\mu)$, where vector
$u_V(\mu)$ is called the reduced solution.  The size of the reduced
solution is the number of snapshots.

Using $V u_V(\mu) \approx u(\mu)$, we can rewrite
Equation~\ref{align:full-problem} as $A(\mu) V u_V(\mu) \approx f(\mu)$.
This is an overdetermined system.  Therefore, we multiply the full
problem from left with $V^T$, which yields our reduced problem $V^T
A(\mu) V u_V(\mu) = V^T f(\mu)$.  We call $A_V(\mu) := V^T A(\mu) V$ the
reduced matrix with snapshot matrix $V$.  Notice that $A_V(\mu)$ is
again a separable matrix.  The matrices ($V^TA_iV$) in this separation
can be pre-computed, and the matrix $A_V(\mu)$ can be rapidly assembled:
\begin{align*}
  \mathbf{V}^T \mathbf{A}(\mu) \mathbf{V} & =
  \mathbf{V}^T \left ( \theta_1 \left (\mu \right )\mathbf{A}_1 + \cdots
  + \theta_m(\mu) \mathbf{A}_m \right ) \mathbf{V} \\
  & = \theta_1(\mu) \mathbf{V}^T \mathbf{A}_1 \mathbf{V} + \cdots +
  \theta_m(\mu) \mathbf{V}^T \mathbf{A}_m \mathbf{V}
\end{align*}
Similarly, $f_V(\mu) := V^T f(\mu)$ can be partially pre-computed and
rapidly assembled.

The process of computing a reduced solution is now to solve $A_V(\mu)
u_V(\mu) = f_V(\mu)$ and then to reconstruct $u_V(\mu)$ to the full
problem space by multiplication with $V$.  Solving the low dimensional
problem is much faster than solving the full problem, as the low
dimensional problem has only the size of the number of snapshots, which
is typically much smaller than the full problem size.

Using a reduced basis instead of solving the full problem typically
degrades the quality of the solution.  To express the quality of the
solution, we use the residual norm of the approximation as error
indicator.  The residual is the difference of $A(\mu) V u_V(\mu)$ and
the right-hand side $f(\mu)$.  If the residual $r$ is $0$, the
approximation $V u_V(\mu)$ is the exact solution $u(\mu)$ of the
algebraic problem $A(\mu) u(\mu) = f(\mu)$.  The residual is a
well-known measurement for approximations in numerics and can be
computed very efficiently using pre-computation (c.f. Appendix). This
allows for fast quality checks for specific parameters during run-time.

\subsection{The Basis Generation Process}

\noindent
Using the residual as error indicator, the snapshots can be computed
from a training set of parameters using a \emph{greedy}
approach~\cite{Veroy2003}.

\begin{figure}[bth]
  \begin{algorithmic}[1]
    \Function{GreedyBasisGeneration}{$\mathit{train\_set}, max\_res$}
      \State $\mathit{basis} \gets $ initial basis
      \State $\mathit{residuals}[\mu] \gets
        \mathit{basis}.\mathit{residual}(\mu) \quad \forall \mu \in
        \mathit{train\_set}$
      \While{$\mathsf{max}(\mathit{residuals}) \geq \mathit{max\_res}$}
        \State $\mu^*\gets \mathsf{max}(\mathit{residuals}).\mathit{key}$
        \State $\mathit{solution} \gets $ solution of the full problem for
        $\mu^*$
        \State add $\mathit{solution}$ to $\mathit{basis}$
        \State $\mathit{residuals}[\mu] \gets
          \mathit{basis}.\mathit{residual}(\mu) \quad \forall \mu \in
          \mathit{train\_set}$
      \EndWhile
      \State \Return $\mathit{basis}$
    \EndFunction
  \end{algorithmic}
  \caption{%
    Greedy approach for generation of a reduced basis, where
    $\mathit{train\_set}$ is a set of parameters and $\mathit{max\_res}$
    is the maximum residual threshold.
  }
  \label{fig:greedy}
\end{figure}

Figure~\ref{fig:greedy} depicts the pseudo code of the greedy basis
generation method.  The user provides a set of training parameters
($\mathit{train\_set}$) and a maximum threshold for the residual
($\mathit{max\_res}$).  The initial basis can be either an existing
basis, or the reduced basis based on the evaluation of a random training
parameter.  The algorithm will terminate when the generated basis
provides a residual norm lower than the provided threshold for all
parameters in the training set.  In every iteration of the loop, one
solution of the full problem is computed and added to the basis.  For
this computation, the parameter that yields the maximum residual norm
using the current basis is chosen.

\subsection{Limitations of the Reduced Basis Method}

\noindent
The RBM converts the full dimensional problem into a lower dimensional
problem by computation on snapshots.  The dimension of the reduced
problem depends on the number of snapshots.  The number of snapshots
depends on the characteristics of the problem and the quality required
by the user.  The here introduced method works only for stationary
problems without any changes to the geometry.  However, in recent
literature, RBM approaches for time-dependent problems or problems with
changing geometry have already been proposed~\cite{Haasdonk2014,
Rozza2008}.  While such approaches could be used in the approaches
presented in this article, we will focus on stationary problems and
fixed-geometry.


\section{System Model}
\label{sec:model}

\begin{figure}[htb]
  \centering
  \begin{tikzpicture}[node distance=4.5em]
    \node [rectangle, draw=black, thick, align=center]
      (middleware)
      {\footnotesize\textbf{Mobile Simulation} \\ \footnotesize\textbf{Middleware}};
    \node [rectangle, draw=black, thick, above of=middleware,
    xshift=-5em, text width=5em, align=center]
      (num)
      {\footnotesize Numerical \\ Simulation};
    \node [rectangle, draw=black, thick, above of=middleware,
    xshift=5em, text width=5em, align=center]
      (app)
      {\footnotesize User \\ Application};
    \node [rectangle, draw=black, thick, below of=middleware,
    xshift=5em]
      (m)
      {\footnotesize Mobile};
    \node [rectangle, draw=black, thick, below of=middleware,
      xshift=-5em]
      (s)
      {\footnotesize Server};
    \node [right of=s, yshift=0.4cm, node distance=5em] (wireless) {
      \includegraphics[width=0.8cm]{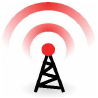}
    };
    \draw [<->, ultra thick, color=blue!40] (s) -- (m);
    \draw [<->, thick] (num) --
      node[below, midway, yshift=0.5em, xshift=-2.5em]
      {\footnotesize snapshots}
      (middleware);
    \draw [<->, thick] (app) -- 
      node[below, midway, yshift=0.5em, xshift=2em]
      {\footnotesize queries}
      (middleware);
    \draw [<->, thick] (m) -- (middleware);
    \draw [<->, thick] (s) -- (middleware);
  \end{tikzpicture}
  \caption{System Model}
  \label{fig:system-model}
\end{figure}
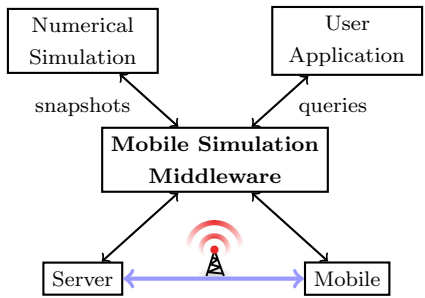

\noindent
Next, we describe the assumptions on hardware and software components,
as well as the interfaces between the components in our mobile
simulation middleware.  Figure~\ref{fig:system-model} depicts an
overview of the system components and interfaces.

\subsection{System Components}

\noindent
The system consists of two compute nodes, namely the mobile device and
the server.  Both nodes are connected via a wireless communication
channel.  Furthermore, the system consists of two software components
provided by the application programmers, the numerical simulation and
the user application, and the middleware, which defines the distribution
of computations.

The mobile device is the augmented reality headset carried by the user.
Energy consumption on the mobile device is critical as it is
battery-powered. There are two distinct energy consumers on the mobile
device, the processor and the communication module.

In contrast to mobile devices, the server provides fast execution.  It
can be scaled-up by using specialized hardware, such as GPUs for
efficient computation of numerical codes, or scaled-out by adding more
servers in a cloud infrastructure.

For the wireless communication channel between mobile device and server,
we assume data rates of multiple Mbit/s, as provided by state of the art
wireless communication technologies like IEEE 802.11 (WiFi) or 4G
cellular networks.

The numerical simulation is implemented by the simulation expert.  The
simulation problem is implemented as a separable matrix $A(\mu)$ and a
separable vector $f(\mu)$ representing the simulation problem $A(\mu)
\cdot u(\mu) = f(\mu)$ as described in Section~\ref{sec:rbm}.
Parameters of the simulation model are represented by a vector $\mu$.
Additionally, the simulation expert has to define the quality
requirements of the application.  The quality has to be specified by two
parameters.  The first parameter, say $\mathcal{D}$, is the
discretization of the full problem.  The second, say $r_{\textit{max}}$,
is the maximum residual value, which is an indicator for the error
introduced by the RBM.

The user application is implemented by the application programmer.  It
sends queries to the middleware.  Queries contain a parameter vector
$\mu$, which encodes sensor data or user input.  When the query is
answered, the user application visualizes the simulation results on the
augmented reality headset.

The mobile simulation middleware connects the components.  It executes
code on the server and on the mobile device.  Intuitively, the reduced
basis method will be used to answer queries with low latency on the
mobile device, and the compute-intensive pre-computation of the reduced
basis will be performed on the server.

\subsection{Interfaces}

\begin{figure}[t]
  \centering
  \begin{tikzpicture}[node distance=10em]
    \node [rectangle, draw=black, thick, align=center, rotate=90, text
    width=6em]
      (middleware)
      {\footnotesize\textbf{Mobile} \\
      \footnotesize\textbf{Simulation} \\
      \footnotesize \textbf{Middleware}};
    \node [rectangle, draw=black, thick, align=center, rotate=90,
      above of=middleware, text width=6em]
      (simulation)
      {\footnotesize Numerical \\ \footnotesize Simulation};
    \node [rectangle, draw=black, thick, align=center, rotate=90,
      below of=middleware, text width=6em]
      (application)
      {\footnotesize User \\ \footnotesize Application};
    \draw [->, thick] ([yshift=2em] middleware.north) --
    node[above, midway, yshift=-0.2em, xshift=0em]{\footnotesize problemMatrix()}
    ([yshift=2em] simulation.south);
    \draw [->, thick] ([yshift=0em] middleware.north) --
    node[above, midway, yshift=-0.2em, xshift=0em]{\footnotesize rightHandSide()}
    ([yshift=0em] simulation.south);
    \draw [->, thick] ([yshift=-2em] middleware.north) --
    node[above, midway, yshift=-0.2em, xshift=0em]{\footnotesize snapshot($\mu$)}
    ([yshift=-2em] simulation.south);
    \draw [->, thick] (application) --
    node[above, midway, yshift=0em, xshift=0em]{\footnotesize handleQuery($\mu$)}
    (middleware);
  \end{tikzpicture}
  \caption{Interfaces for the mobile simulation middleware.}
  \label{fig:interfaces}
\end{figure}
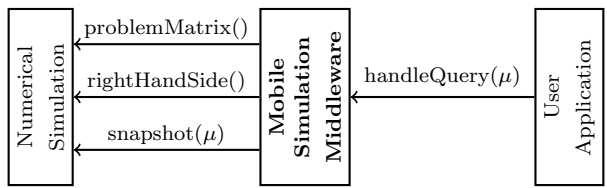

\noindent
The numerical simulation and the user application provide interfaces for
the mobile simulation middleware.  Figure~\ref{fig:interfaces} shows an
overview of all interfaces.  There are three methods of the numerical
simulation to be called by the middleware and one method called by the
user application.

The numerical simulation has to implement three interfaces providing the
problem matrix, the right-hand side, and solutions to the simulation
problem.  The problem matrix and the right-hand side has to be provided
in parameter separable form.  This call is only depending on the quality
parameter $\mathcal{D}$.  The interface to provide solutions of the
simulation problem, called \emph{snapshot}, provides $u(\mu)$ as the
solution of the problem $A(\mu) u(\mu) = f(\mu)$ depending on the
parameter.  Notice that the implementation of the interface to provide
snapshots is optional.  The mobile simulation middleware could also use
a generic algorithm to solve this problem. However, the simulation
expert usually knows which solver should be used to solve the simulation
problem efficiently.

The user application sends queries to the mobile simulation middleware.
Queries contain the parameter $\mu$.  The middleware will return an
approximate solution which fulfills the quality requirements given by
the simulation expert.


\section{Basic Approach}
\label{sec:basic}

\noindent
In the following, we present our approaches for the efficient execution
of mobile simulations using the Reduced Basis Method.  We first present
a basic approach in this section, which is then further extended to
improve adaptability and energy efficiency in the following sections.

The basic approach for processing queries with different parameters on
the mobile device consists of four steps:
\begin{inparaenum}[(1)]
  \item generation of the reduced basis on the server;
  \item communication of the reduced basis from the server to the mobile
    device where the reduced basis is stored on the internal storage;
  \item loading the reduced basis from the internal storage of the
    mobile device;
  \item processing queries on the mobile device using the reduced basis.
\end{inparaenum}

The generation of the basis is executed on the server.  To this end, the
mobile device sends a request to the server which contains all
information needed for the basis generation process.  This includes the
training set and the minimum quality as maximum residual threshold,
which depends on the application.  The training set can be given by the
domain expert or, in applications where sensor values are read, the
mobile device can first collect some sensor data, statistically obtain
the distribution of the parameter $\mu$, and then use this distribution
to create the training set for the reduced basis.

Once the basis has been generated on the server, it is sent to the
mobile device.  The mobile device stores the basis on internal storage.
Notice that the pre-computation of the reduced basis can take multiple
minutes, depending on the numerical simulation code, the training set,
and the number of snapshots needed to achieve the quality as specified.
However, this step is only needed once for initialization and should not
be performed when latency-sensitive queries need to be processed.

\begin{figure}[h]
  \centering
  \begin{tabular}{l l}
    \textbf{Data} & \textbf{Size} \\
    \hline
    Snapshots & $n\cdot d$ \\
    Reduced Problem Matrix & $S_A \cdot n^2$ \\
    Reduced Right Hand Side & $S_f \cdot n$ \\
    Residual Computation Matrices &
      $S_{A}^2 n^2 + 2 S_{A} S_{f} n + S_{f}^2$
  \end{tabular}
  \caption{Size of the reduced basis in floating point numbers.}
  \label{fig:tbl-size}
\end{figure}

Figure~\ref{fig:tbl-size} lists the size of the data communicated and
stored on the mobile device.  The size of the data depends on the number
of snapshots $n$, the number of discretization points of the full
problem $d$, and the number of summands in the separation of the problem
matrix $S_A$ and the right-hand side $S_f$.  The number of
discretization points $d$, which depends on $\mathcal{D}$, is by far the
largest part, typically multiple thousand floating point numbers (in our
evaluation up to $65536$ with $\mathcal{D} = 256$). The number of
snapshots depends on the residual and is typically below $30$ in our
experiments.  Numbers $S_A$ and $S_f$ are constant for a given problem.
In our evaluation these values were $4$ and $1$.

After the basis is stored in a file on the mobile device, this file
needs to be read by the middleware on the mobile device.  As the file
size for the reduced basis can grow rapidly, reading the data from the
file can take up to seconds.  However, this step is needed only once and
can be performed when the user starts the user application, long before
the first query will be received by the middleware.  The basis can then
be stored in memory for processing of multiple queries.

Processing a query is then straightforward as described in
Section~\ref{ssec:reduced-problem}.  First, we need to assemble the
reduced system and then compute the solution of the reduced problem.
After that, we need to multiply the solution with the snapshots to get
an approximate solution of the full problem.  In addition to the
approximate solution, we also calculate the residual of this
approximation and provide this information to the application.  Notice
that fulfilling the quality constraints for queries with parameters
outside the range of the training set cannot be guaranteed using this
approach.  However, it is known that the quality of the result does
depend on the region of the parameters rather than the density or
specific choice of parameters in the training set~\cite{Haasdonk2014}.
Therefore, for queries with parameters inside the range of the training
set, the resulting approximation should have high quality.  Furthermore,
for many practical problems, the parameter region is known a priori by
physical constraints.  For example, if one parameter is the heat
conductivity of some material, the application can request the reduced
basis in the range of all materials to be used for the specific purpose,
e.g., all exhaust tubes ever used by the company.

The basic approach has several drawbacks.  First, the parameter range
needs to be known before the basis generation process.  If the parameter
range changes, e.g., because the range of sensor values changes, the
approach has to start from scratch.  We therefore present an adaptive
approach in the next section.  Another problem is the latency and energy
overhead introduced by reading the reduced file from internal storage of
the mobile device.  This is significantly improved using the subspace
approach, which will be presented in Section~\ref{sec:subspace}.


\section{Adaptive Approach}
\label{sec:adaptive}
\graphicspath{ {figs/adaptive-approach/} }

\noindent
If the parameter range and distribution are not known a priori, the
basic approach might not be able to fulfill the constraint on quality
for all queries.  We therefore introduce an adaptive approach next that
refines the basis during runtime.  This approach is more flexible and
also suitable for harder simulation problems, i.e., problems that need
more snapshots to fulfill the user requirements.

\subsection{Overview}

\noindent
The adaptive approach builds upon the basic approach.  Similar to the
basic approach, some initial reduced basis is made available on the
mobile device as described in the previous section.  However, in
contrast to the basic approach, when a new query $q$ arrives, the
adaptive approach first computes the residual of the approximate
solution provided by the RBM.  If the residual fulfills the quality
requirements of the application, the query will be answered with the
approximate solution.  If the residual does not fulfill the requirements
of the application, the mobile device will request an update of the
reduced basis from the server.  Once the mobile device receives the
update, it can again compute the approximate solution, which will---as a
property of the RBM---return the exact solution of the full problem.
Figure~\ref{fig:adaptive-approach} depicts the pseudocode of the
adaptive approach.

\begin{figure}[tb]
  \centering
  \begin{algorithmic}[1]
    \Function{onQueryReceived}{$q$}
      \State $\mu \gets$ parameter of request $q$
      \State $\mathit{basis} \gets$ basis available on mobile 
      \If{$\mathit{basis}.\mathit{residual}(\mu) \leq max\_res$}
        \State \Return approximate solution using $\mathit{basis}$
      \EndIf
      \State send $\mu$ to server; receive basis update
      \State apply basis update to $\mathit{basis}$
      \State \Return approximate solution using $\mathit{basis}$
    \EndFunction
  \end{algorithmic}
  \caption{Pseudocode for the adaptive approach.}
  \label{fig:adaptive-approach}
\end{figure}

In the following, we will describe the parts of the approach, including
the computation of the error indicator and content of the
server request, and the processing of the update on the server.

\subsection{Error Indicator and Server Requests}

\noindent
In addition to the basic approach, for handling query $q$, the mobile
device has to compute error indicators for the approximate solution
provided by the RBM.  This error indicator represents the quality of the
approximate solution.  One very generic error indicator is the residual.
The computation of the residual can be implemented very efficiently by
exploiting the parameter separability (c.f. Appendix).

Once the mobile device has computed the error indicator, it can check if
the quality bounds of the user can be met.  If the result is
insufficient, the mobile device will request a basis update from the
server.  This basis update contains the parameter $\mu$ of the query and
the identifier of the reduced basis which is currently used.  As an
identifier, the parameters of the snapshots and the discretization of
the underlying numerical simulation can be used.

\subsection{Computation on the Server}

\noindent
When an update request with parameter $\mu$ and an identifier of the
reduced basis is received by the server, the server first loads the
properties of the reduced basis.  It then computes a solution of the
full problem with parameter $\mu$ and the discretization settings of the
reduced basis.  After computation of the full solution, this solution is
orthogonalized to other basis vectors and is normalized to obtain more
robust numerical systems.  The server then computes the updated
separable problem matrix and the separable right-hand side (c.f.
Section~\ref{sec:basic}).  Last, the updated residual matrices are
computed.  All of these operations require high-dimensional and costly
operations.  However, the most time consuming operation is the
computation of the full solution on the server.  Therefore, there is
only little overhead compared to a pure offloading approach, where only
the full problem solution is computed on the server.

\subsection{Basis Updates}

\noindent
Once the server has computed the update of the reduced basis available
on the mobile device, it sends the update back.  The update includes a
snapshot and updates for the separable matrices.  Most entries of the
matrices can be re-used, and the update does only contain one column and
one row vector of the matrices.  Nevertheless, the size of the update
grows linearly with the number of snapshots included in the reduced
basis.  However, for a small number of snapshots, the dominant part is
still the snapshot of the full problem.  Therefore, the overhead to only
communicating the full problem result is very small (for instance only
$1.13\,\%$ for a 2D problem with $256^2$ points, $S_A =4, S_f = 1$, and
$20$ snapshots).


\section{Subspace Approach}
\label{sec:subspace}

\noindent
In our analysis of the basic approach, we found that reading the
snapshots from internal storage is the major energy-consuming part.  We
therefore present in this and the following section approaches for
reducing the number of snapshots needed for query processing on the
mobile device.  In this section, we present the \emph{subspace
approach}, which limits the computation of the problem to a subspace of
the vector space spanned by all snapshots.

\begin{figure}
  \centering
  \begin{tikzpicture}[%
      snapshot/.style = {
        draw,
        minimum height=3em,
        node distance=1em,
        fill=blue!20
      },
      snapshot-unused/.style = {
        snapshot,
        dashed,
        fill=gray,
      },
      examples/.style = {
        matrix,
        below of=approaches,
        node distance=6em,
        column sep=3em,
      },
    ]

    \node (snapshots) {
      \begin{tikzpicture}
        \node[snapshot] (s1) { };
        \node[snapshot, right of=s1] (s2) { };
        \node[snapshot, right of=s2] (s3) { };
        \node[snapshot-unused, right of=s3] (s4) { };
        \node[snapshot-unused, right of=s4] (s5) { };
        \draw[->]
          ([yshift=-0.5em] s1.south) --
          node[below, midway] {\footnotesize $m$}
          ([yshift=-0.5em] s5.south);
      \end{tikzpicture}
    };

    \node [%
      above of=snapshots,
      node distance=3em,
    ] (label-snapshots) {\footnotesize Snapshots};

    \node [%
      draw,
      fill=gray,
      right of=snapshots,
      node distance=12em,
      minimum height=3em,
      minimum width=3em,
      yshift=0.8em,
      dashed,
    ] (reduced-full) { };

    \node [%
      draw,
      fill=blue!20,
      left of=reduced-full,
      node distance=0em,
      minimum height=2.3em,
      minimum width=2.3em,
      yshift=0.35em,
      xshift=-0.35em
    ] { };

    \draw [->]
      ([yshift=-2em] reduced-full.west) -- 
      node[below, midway] {\footnotesize $m$}
      ([yshift=-2em] reduced-full.east);

    \draw [->]
      ([xshift=2em] reduced-full.north) -- 
      node[right, midway] {\footnotesize $m$}
      ([xshift=2em] reduced-full.south);

    \node [
      above of=reduced-full,
      node distance=3em,
      text width=7em,
      align=center,
    ] {\footnotesize Reduced Problem Matrix};
  \end{tikzpicture}

  \caption{Subspace approach chooses $m$ of $n$ avialable snapshots in
    the order given during the basis generation approach and changes the
    reduced problem matrix, the reduced right-hand side, and the
    residual matrices (last two not depicted).
  }
  \label{fig:subspace-approach-m}
\end{figure}
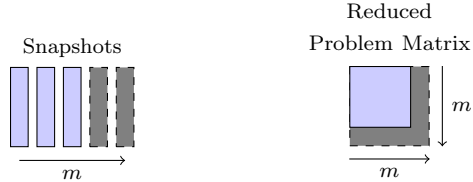

The reduced basis is generated such that it fulfills quality
requirements for all parameters in the training set.  However, for one
specific parameter $\mu$, it might be sufficient to compute on fewer
snapshots.  In the subspace approach, we therefore limit the computation
to the first $m$ snapshots in the order given by the reduced basis.
Therefore, if $n$ snapshots $s_1, \dots, s_n$ are given, we want to find
$m \leq n$ such that the quality constraint is still fulfilled and
compute an approximation only on $s_1, \dots, s_m$ (c.f.
Fig.~\ref{fig:subspace-approach-m}).  This saves us from reading $n - m$
snapshots while still fulfilling the quality requirements of the user.

The subspace approach is divided into two problems.  First, we explain
how we can reuse the data structure of the matrix for computation on a
subspace.  Second, we explain how we can find the snapshot given the
quality constraint by the user.  Last, we shortly discuss how this
approach can be combined with the adaptive approach.

\subsection{Reusing the Data Structure}

\noindent
For computing a solution on the reduced basis spanned only by the
snapshots $s_1, \dots s_m$, we can reuse the existing data structure.
We can compute on sub-matrices which are created when trimming rows from
the right and columns from the bottom.

For the reduced problem matrix, we just need the first $m$ rows and the
first $m$ columns.  Similarly, we only need the first $m$ entries of the
reduced right-hand side.  The residual matrices can be trimmed
similarly.  Notice that the right-hand side and the reduced problem
matrix are separable matrices.  Trimming the separable form of the
matrices therefore includes trimming multiple matrices.

Notice that the reuse of the data structure is essential at this point.
Re-computation of the reduced problem matrix would otherwise involve the
high dimensional problem matrix $A(\mu)$ and the snapshots.  Using 
the sub-matrices, neither the problem matrix, nor the snapshots are
needed for computation of the residual for the subspaces.

\subsection{Subspace Selection}

\noindent
Now that we know how to compute a solution of the reduced problem by
computation on the sub-matrices, we want to find $m$, such that
computing on snapshots $s_1, \dots, s_m$ gives us a solution that
fulfills the quality requirements of the user.  We call the subspace
spanned by the first $m$ snapshots $S(m)$.

In order to find $m$ for $S(m)$, we use a linear search.  When a query
arrives with parameter $\mu$, we first load the reduced problem matrix
and the residual matrices into memory.  We then loop, starting with $m =
n$, compute the subspace $S(m)$, and compute the residual for parameter
$\mu$ on $S(m)$, until we find the lowest $m$ such that $S(m)$ fulfills
the quality requirements.  Once this $m$ is known, we load the $m$
snapshots from the file into memory and reconstruct the reduced solution
in the full problem space.

The linear search could also be bottom-up starting with one snapshot or
could be replaced by a bisection approach.  However, this would result
in longer search time when the number of snapshots needed is high.

The subspace approach can also be used with the adaptive approach.  If
the quality check for $m = n$ fails, the mobile device can request a
basis update from the server.  We then have a three-level storage model,
where snapshots are either stored in-memory, on internal storage, or on
the server.


\section{Reorder Approach}
\label{sec:reorder}

\noindent
For the subspace approach, the order of the snapshots is fixed.  This
might lead to suboptimal solutions, e.g., when the query has the same
parameter as the last snapshot.  In this example, the subspace approach
needs to choose all snapshots.  If we would reorder the snapshots
according to the importance of the snapshots, then the snapshot with the
same parameter would be the first and the subspace with only the first
snapshot would already be sufficient.  This motivates our reorder
approach, which we introduce in this section as a preceding step to the
subspace approach.  The reorder approach operates on pre-computed data
in order to allow the subspace approach to reduce the number of
snapshots needed for the computation.

\begin{figure}
  \centering
  \begin{tikzpicture}[%
      snapshot/.style = {
        draw,
        minimum height=3em,
        node distance=1em,
      },
      snapshot-unused/.style = {
        snapshot,
        dashed,
        color=black!30,
      },
      examples/.style = {
        matrix,
        below of=approaches,
        node distance=6em,
        column sep=3em,
      },
    ]

    \node[matrix, anchor=north, column sep=5em] {

      \node (snapshots) {
          \begin{tikzpicture}

            \node[snapshot, fill=orange] (s1) { };
            \node[snapshot, right of=s1, fill=blue] (s2) { };
            \node[snapshot, right of=s2, fill=red] (s3) { };
            \node[snapshot, right of=s3, fill=green] (s4) { };
            \node[snapshot, right of=s4, fill=yellow] (s5) { };

          \end{tikzpicture}
      }; &

      \node[] (reorder) {
        \begin{tikzpicture}

          \node[snapshot, fill=green] (s1) { };
          \node[snapshot, right of=s1, fill=yellow] (s2) { };
          \node[snapshot, right of=s2, fill=blue] (s3) { };
          \node[snapshot, right of=s3, fill=orange] (s4) { };
          \node[snapshot, right of=s4, fill=red] (s5) { };

          \path[->]
            (s1.south) edge [<->, bend right=90] node {} (s4.south)
            (s2.south) edge [bend right=90] node {} (s3.south)
            (s3.south) edge [bend right=90] node {} (s5.south)
            (s5.south) edge [bend left=90] node {} (s2.south)
          ;

        \end{tikzpicture}
      }; &

      \node[] (subspace) {
        \begin{tikzpicture}

          \node[snapshot, fill=green] (s1) { };
          \node[snapshot, right of=s1, fill=yellow] (s2) { };
          \node[snapshot, right of=s2, fill=blue] (s3) { };
          \node[snapshot, dashed, right of=s3, fill=orange!70] (s4) { };
          \node[snapshot, dashed, right of=s4, fill=red!70] (s5) { };

          \draw[->]
            ([yshift=-.5em] s1.south) --
            node[below, midway] {\footnotesize $m$}
            ([yshift=-.5em] s5.south);

        \end{tikzpicture}
      }; \\
    };

    \node [below of=snapshots, node distance=2.2em] {\footnotesize Snapshots};

    \draw [->, thick] ([yshift=-2em] snapshots.north east) --
      node [above, midway] {\footnotesize 1. Reorder} 
      ([yshift=-2em] reorder.north west);

    \draw [->, thick] ([yshift=-2em] reorder.north east) --
      node [above, midway] {\footnotesize 2. Subspace}
      ([yshift=-2em] subspace.north west);
  \end{tikzpicture}
  \caption{%
    The reorder approach permutes the snapshots before choosing subspace
    with $m$ snapshots depending on parameter $\mu$.  
  }
  \label{fig:reorder-approach}
\end{figure}
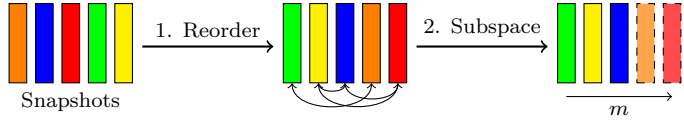

Figure~\ref{fig:reorder-approach} depicts the idea.  First snapshots are
reordered.  Then a subspace is chosen using the previously introduced
subspace approach.  The reordering depends on the query parameter $\mu$.
As we will show, finding a reordering can be implemented on the
pre-computed data such that it can be executed efficiently and fast on
the mobile device.

There are two problems for reordering snapshots:
\begin{inparaenum}[(1)]
  \item find a suitable reordering for parameter $\mu$, and
  \item perform the reordering by re-using pre-computed data.
\end{inparaenum}
To improve latency and energy cost, we need to find a good order of the
$n!$ possible orders in the first step and solve the second problem by
only using pre-computed data and not require any high-dimensional
operations of the numerical simulation.

\subsection{Finding an Order}

\noindent
As the subspace approach will compute an approximate solution using the
first $m$ snapshots, we want to minimize the difference from $m=n$
snapshots to $m=n-1$ snapshots.  If $u$ is the reduced solution for
subspace $m=n$ and $\tilde{u}$ is the reduced solution for subspace
$m=n-1$, the difference in the approximate solutions after the
reconstruction is
\begin{align}
  \left \lVert
  \sum_{i=1}^n \mathbf{u}^{(i)} \mathbf{s}_i -
  \sum_{i=1}^{n-1} \mathbf{\tilde{u}}^{(i)} \mathbf{s}_i
  \right \rVert
  \leq
  \left |u^{(n)} \right| +
  \left \lVert
  \sum_{i=1}^{n-1} \mathbf{s}_i
  \left (
    \mathbf{u}^{(i)} - \mathbf{\tilde{u}}^{(i)}
  \right)
  \right \rVert,
\end{align}
where $u^{(i)}$ is the $i$-th entry in vector $u$ and we assume that
snapshots are normalized.  This motivates to move the snapshots with
lowest absolute coefficient to the end.  We therefore order the
snapshots according to the absolute value of their reduced solution in
descending order.  Notice that this step does not need the
reconstruction of the approximation and therefore no snapshots, but only
the reduced problem in memory.

\begin{figure}[bth]
  \centering
  \begin{algorithmic}[1]
    \Function{findReorder}{$\mu$}
      \State $\left(u_V^{(i)}(\mu)\right )_{i=1}^n \gets$ coefficients
        of the reduced solution for parameter $\mu$
        \State $t \gets \left\{ \left (|u_V^{(i)}(\mu)|, i\right ) \right\}_{i=1}^n$
      \State sort $t$ using the first element of the tuples
      \State \Return order of snapshots as second elements in $t$
    \EndFunction
  \end{algorithmic}
  \caption{Finding reordering for normal bases}
  \label{fig:find-reorder}
\end{figure}

Figure~\ref{fig:find-reorder} depicts the pseudo code for finding the
reordering.  We need the pre-computed reduced problem in memory, which
consists of the reduced problem matrix $A_V$ and the reduced right-hand
side $f_V$.  This are separable matrices which do depend on the snapshot
matrix $V$ (c.f.  Section~\ref{ssec:reduced-problem}).  We first compute
the coefficients as solution $u_V(\mu)$ of the reduced problem $A_V(\mu)
u_V(\mu) = f_V(\mu)$ and sort them according to their absolute value.
The reordering is represented as a list, where the $j$-th position has
value $i$ when the $j$-th snapshot should be moved to position $i$.

\subsection{Reordering Precomputed Data}

\noindent
The reordering can be represented as a permutation matrix $P$.
Reordering can then be executed by multiplying the existing snapshot
matrix $V$ with the permutation matrix $P$.  Using this approach, we can
reuse the pre-computed data, such as the separable reduced problem
matrix, the separable right-hand side.

The reordering can be applied to the separable reduced problem matrix by
permuting rows and columns.  We use the pre-computed data for the
snapshot matrix $V$ and show that we can reorder this data in order to
obtain the pre-computed data for matrix $V' = VP$, where $P$ is the
permutation matrix of our reordering.  The available pre-computed data
is $A_V(\mu) := V^T A(\mu) V$.  We can reuse this data for the reordered
snapshots $V'$, since $A_{V'}(\mu) = (VP)^T A(\mu) (VP) = P^T V^T A(\mu)
V P = P^T A_V(\mu) P$.  To obtain the reduced problem matrix,
we only need to multiply with permutation matrix $P^T$ from left, which
is a permutation of the columns, and permutation matrix $P$ from right,
which is a permutation of the rows.

The separable right-hand side and the residual matrices can be reordered
by ordering of the entries in the vectors analogously to the reduced
problem matrix.

Notice, that the permutation of the matrix can be implemented much
faster than multiplication with the permutation matrix by reordering
rows and columns in the underlying data structure of matrices and
vectors.  However, we used the notation for multiplication to show the
correctness of the approach.

To improve numerical stability, snapshots can be orthonormalized for the
previous approaches.  However, in this approach, it is beneficial to
only normalize the snapshots.  Orthogonalization, e.g. by using
Gram-Schmidt, reinforces the order and reduces the flexibility of the
reorder approach.


\section{Reorder Basis Generation}
\label{sec:reorder-basis-generation}

\noindent
In order to optimize the number of snapshots to be used in the reorder
approach, we next present a new approach for basis generation.  This
approach takes into account the online computation using the reorder
approach and is therefore able to generate better reduced bases on the
server.  To this end, we modify the previously introduced greedy basis
generation (cf. Sec.~\ref{sec:rbm}) to allow for better performance
for the reorder approach.  

In contrast to the greedy basis generation approach, we define the
$m$-residual as the residual after the execution of the reorder approach
and choosing only $m$ snapshots (see Fig.~\ref{fig:reorder-residual}).
This way, we use the results as provided by our approach already during
the construction of the reduced basis.

\begin{figure}
  \centering
  \begin{algorithmic}[1]
    \Function{ReorderResidual}{$\textit{basis}, l, \mu$}
      \State $r \gets \textsc{FindReorder}(\textit{basis}, \mu)$
      \State $b \gets \textsc{ApplyReorder}(\textit{basis}, r)$
      \State $s \gets \textsc{CutBasis}(b, l)$
      \Comment{As in subspace approach}
      \State \Return $s.\mathit{residual}(\mu)$
    \EndFunction

    \Function{ReorderBasisGeneration}{$T$, $a$, $\mathit{max\_res}$}
      \State $s \gets$ solution of full problem for random $\mu \in T$
      \State $\mathit{snapshots} \gets \emptyset$
      \State $\mathit{basis} \gets$ reduced basis from
        $\mathit{snapshots}$
        \While{$\exists \mu\in T: \textsc{ReorderResidual}(\mathit{basis},
        |\mathit{snapshots}| - a, \mu) > \mathit{max\_res}$}
        \State $\mu^* \gets \max($residuals$).$key
        \State $s \gets$ solution of the full problem for $\mu^*$
        \State $\mathit{snapshots} \gets \mathit{snapshots} \cup \{s / \lVert s\rVert\}$
        \State compute new $\mathit{basis}$ from $\mathit{snapshots}$
      \EndWhile
      \State \Return basis
    \EndFunction
  \end{algorithmic}
  \caption{%
    Pseudocode for reorder basis generation for training set $T$, number
    of additional snapshots $a$, and maximum residual
    $\mathit{max\_res}$.
  }
  \label{fig:reorder-residual}
\end{figure}

However, this approach for optimizing for a subset of the full solution
space introduces some numerical problems as snapshots are not strictly
linear independent.  We therefore use the Moore-Penrose pseudoinverse to
compute the solution.  The pseudoinverse provides a least square
solution, which can even be used when snapshots are linear dependent.

Figure~\ref{fig:reorder-residual} depicts the basis generation procedure
for the reorder approach.  In contrast to greedy basis generation, we
consider the residual after the reorder computation.  We omit a number
of snapshots for the decision which parameter to use for the next basis
refinement.  The snapshots will be first sorted and then cut to simulate
the reorder approach.  However, the number of snapshots to be omitted
should not be too large to avoid overfitting.  Parameter $a$, which
defines how many snapshots should be omitted, is problem dependent. In
preliminary tests, we found good solutions with $a=3$.

Our original approach was to cut the basis to a constant number of
snapshots.  However, this approach leads to overfitting to the training
set and therefore to very large reduced bases that only provides good
results for the training set but not for test queries.  Having a
constant offset to the greedy basis approach avoids overfitting and
produces only slightly bigger bases which enables the reorder approach
to have more choices when deciding on a reordering.


\section{Evaluation}
\label{sec:evaluation}
\graphicspath{ {figs/evaluation/} }

\noindent
In this section, we present the evaluation of our five approaches using
the Reduced Basis Method (RBM) with respect to energy efficiency and
execution time.  For comparison, we also implemented two simple
solutions without the RBM, the \emph{server-only} and the
\emph{mobile-only} approach.  The server-only approach sends the
combination of parameters to the server.  The solution of the full
simulation problem is computed on the server and sent back to the
mobile.  The mobile-only approach computes the full simulation problem
on the mobile device.

In our evaluation, we consider different performance metrics.  First, we
evaluate the quality of reduced bases with different sizes and compare
different basis generation methods.  Then, we evaluate the runtime and
energy consumption for two workloads, single queries and multiple
queries.

\subsection{Evaluation Setup}

\noindent
Before we present the evaluation results, we introduce our evaluation
setup.  The setup consists of two different mobile devices, a mobile
network, and the setup for measuring the energy consumption.  We also
provide details about the used libraries in the implementation and the
simulation problem used for the evaluation.

Two different devices were used for the runtime evaluation, a Samsung
Note 4 (SM-N910F) and a Samsung Galaxy S7 (SM-G930F).  Both devices use
the Android platform (version 6.0) based on the Linux kernel (version
3.10).  The results for both devices were very similar for most
evaluations.  Therefore, if not stated otherwise, results of the
approaches presented in this section are taken using the Note 4.

For wireless communication, we used IEEE 802.11 (WiFi).  Using
\texttt{ping}, the measured latency between mobile device and server was
between 1.4\,ms and 6.7\,ms with an average of 3.9\,ms.  \texttt{iPerf}
measured bandwidths between mobile and server between 55\,MBits/s and
74\,MBits/s.

\begin{figure}
  \centering
  \begin{subfigure}{.49\linewidth}
    \centering
    \includegraphics[height=0.86\columnwidth,angle=90]{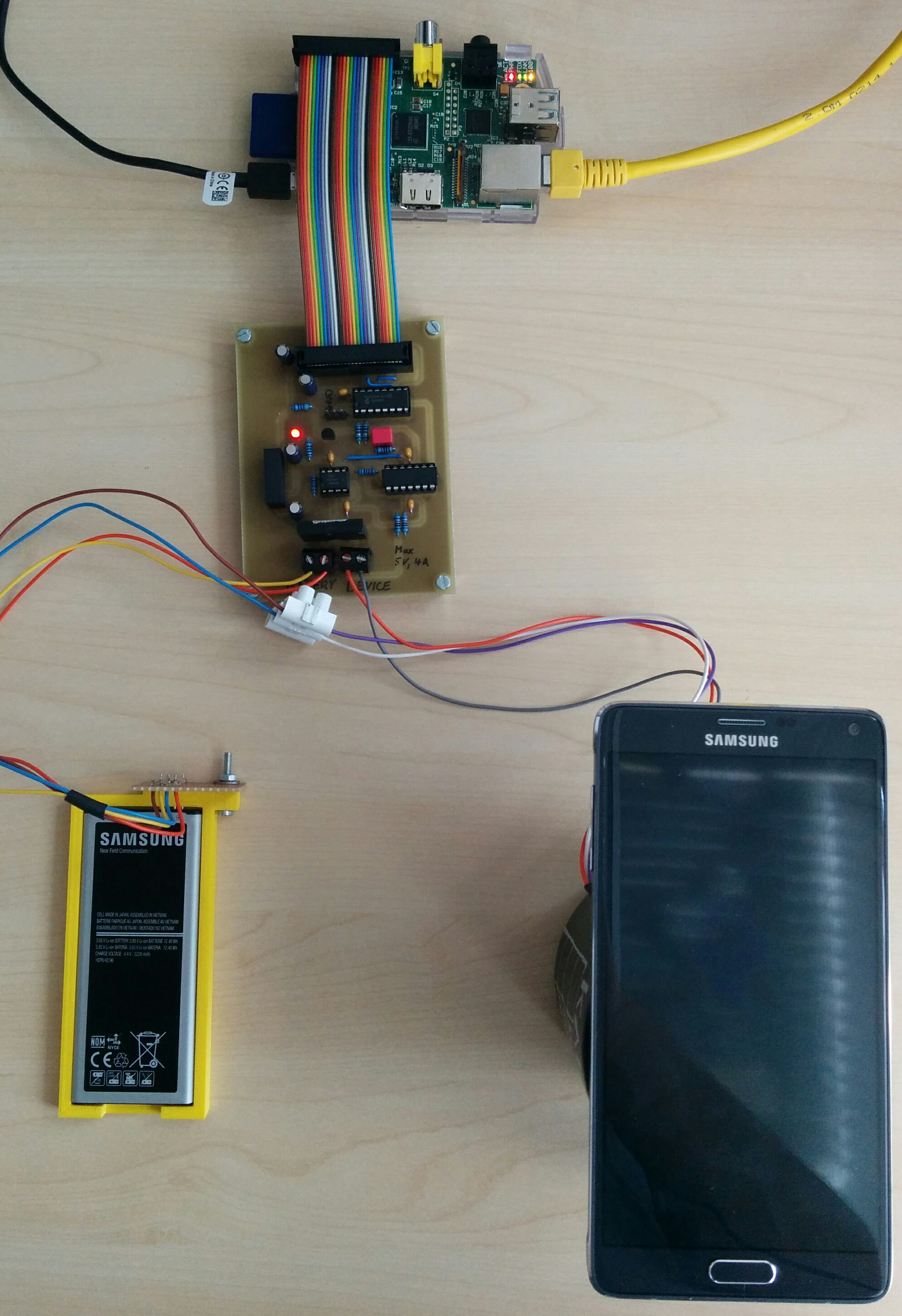}
    \caption{Equipment for measuring energy consumption.}
    \label{fig:equipment}
  \end{subfigure}\hfill
  \begin{subfigure}{0.49\linewidth}
    \centering
    \includegraphics[width=\columnwidth]{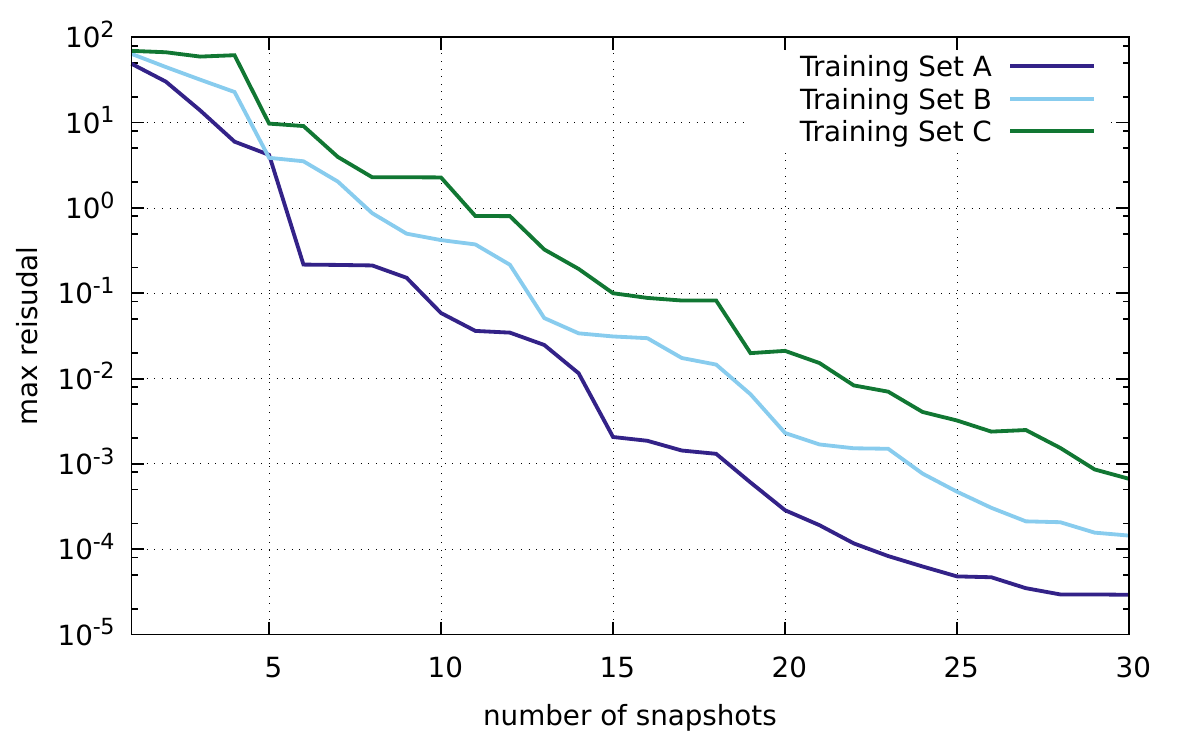}
    \caption{%
      Quality of the RBM with different number of snapshots.
    }
    \label{fig:snapshots-quality}
  \end{subfigure}
  \caption{}
  \label{fig:equipment-snapshots}
\end{figure}

For the energy measurements, we used a custom measurement board with
analog-to-digital converter connected to a Raspberry
Pi%
\footnote{%
  available at \url{https://github.com/duerrfk/rpi-powermeter}
}.
We designed a battery holder and battery replacement in order to perform
energy measurements in situ.  Figure~\ref{fig:equipment} shows our
energy evaluation setup.  All energy measurement values in this section
are absolute values taken from the Note 4 with fixed screen
brightness.  The power consumption of the device in idle mode was
0.4\,W.

As simulation problem for the evaluation, we used the stationary
diffusion-advection equation.  This equation can be used to simulate the
heat in an object as for the application for placing a hot tube as
mentioned in the introduction.  The equation has three parameters, one
for the diffusion ($\mu_{\textit{diff}}$) and two for advection
($\mu_{\textit{advx}}$ and $\mu_{\textit{advy}}$).  For the
implementation, we discretized the equation using finite differences.
As numerics library, we used the Apache Commons Maths library (version
3.6), which is the most popular Java numerics library, and NumPy
(1.13.0) and SciPy (0.19.1), which provide hardware optimization on the
server.  Additionally to the pure Java implementation, we implemented
the mobile-only approach natively using the Android Native Development
Kit (NDK) and the Eigen C++ library in version 3.2.8.

\subsection{Basis Generation Methods}

\noindent
Next, we evaluate the performance of the basis generation methods and
how many snapshots are needed during query processing on the mobile
device.

In order to quantify the number of snapshots needed to reach a certain
quality, we created three different training sets $A$, $B$, and $C$.
The training sets were chosen such that $A\subset B\subset C$, where
parameters $\mu_{\textit{advx}}$ and $\mu_{\textit{advy}}$ spanned
different parts of the parameter space expressing different behavior of
the model ($[0, 40]^2$ for $A$, $[-40, 40]\times [0,40]$ for $B$,
$[-40,40]^2$ for $C$).  Parameter $\mu_{\textit{diff}}$ was for all
three training sets in $[10, 20]$.  All intervals were discretized with
step width $1.0$.

To quantify the relation between the size of the reduced basis and
quality, we used the three training sets, executed the greedy basis
generation algorithm (c.f.~Section~\ref{sec:rbm}), and recorded the
maximum residual of test sets.  The test sets $A_{t}, B_{t}, C_{t}$
consist of $1000$ random points in the range of $A, B, C$.  The
discretization of the full problem was $256 \times 256$, i.e. $256$
points in $x$ and $y$ direction.

In Fig.~\ref{fig:snapshots-quality} depicts the relation between number
of snapshots and quality for each training set.  Training set $A$, which
has smallest variation in the parameters, has the lowest maximum
residual for a fixed number of snapshots.  Notice that the residuum is
measured in the Euclidean norm. To get a better impression, this norm is
always bigger than the maximum absolute difference of any two points in
the resulting vector. Therefore, a residual of, say $0.1$, means that
the result multiplied by the problem matrix results in a vector which
differs in all $256^2$ entries at most $0.1$ from the right-hand side.
Therefore, for many practical applications, a basis with $10$, $15$, or
$20$ snapshots would provide sufficient quality for this problem.

After evaluating the greedy basis generation and the basic approach, we
now quantify the number of snapshots needed during online computation
for different basis generation methods.  The number of snapshots needed
for the subspace approach and the reorder approach is dynamically
depending on parameter $\mu$.  We also want to quantify the fraction of
snapshots typically needed for these approaches compared to the basic
approach, which always uses all snapshots available in the reduced
basis.

\begin{figure}
  \centering
  \includegraphics[width=0.8\linewidth]{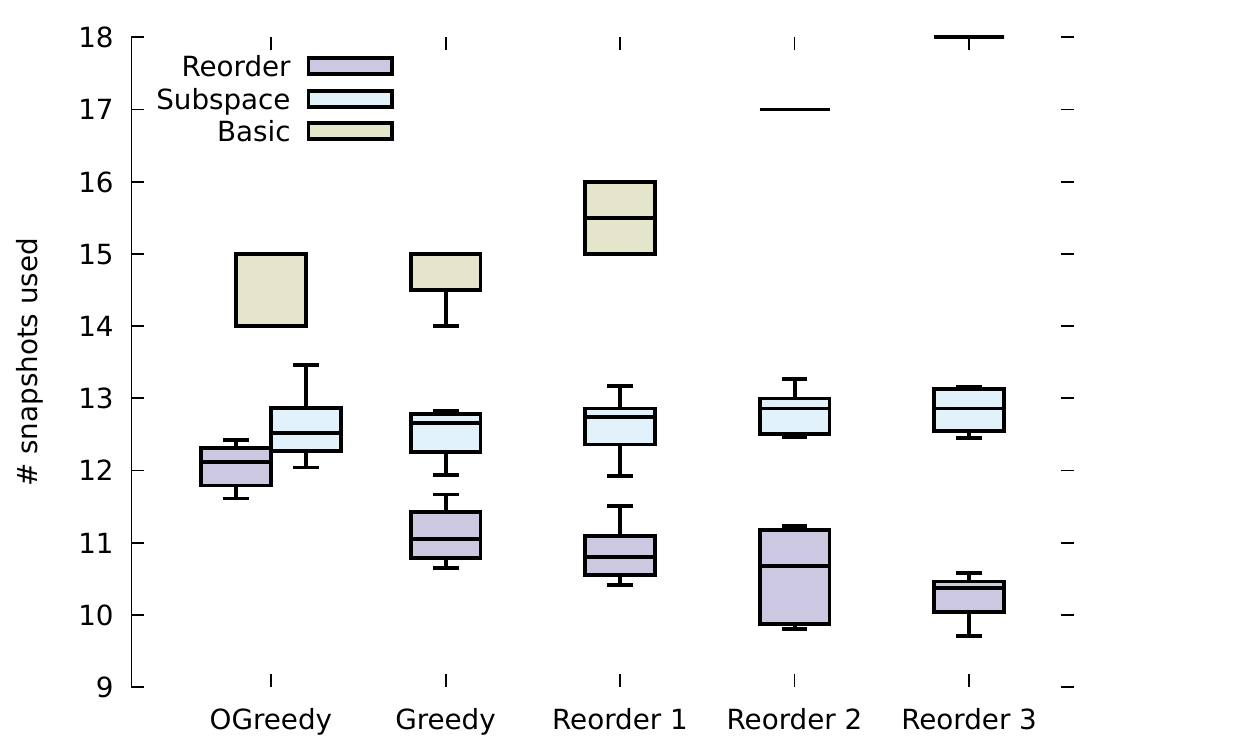}
  \caption{%
    Mean number of snapshots needed in different approaches for
    different reduced basis generation methods.  OGreedy is the
    orthogonal greedy basis.  Reorder $k$ is the reorder basis
    generation method with $k$ additional snapshots.
  }
  \label{fig:reorder-subspace}
\end{figure}

Figure~\ref{fig:reorder-subspace} depicts the number of snapshots needed
for the reorder and the subspace approach in multiple basis generation
runs.  It also compares the greedy basis generation approach to our
reorder basis generation approach.  Using the reorder basis generation
approach, the reorder approach performs better than using the greedy
bases.  In particular, for the reorder basis with $3$ additional the
median mean number of snapshots for the reorder approach is 69.1\,\%
lower than for the basic approach using an orthonormal greedy basis.  On
an orthonormal basis, the median mean number of snapshots of the
subspace approach is only 83.5\,\% compared to the basic approach.
Notice that for the reorder basis generation, the bigger $k$ is in the
additional number of snapshots, the harder it is to generate the reduced
basis because of numerical instabilities.

As these results suggest, we assume in the following that the subspace
approach will only need 83.5\,\% of the snapshots and the reorder
approach only 69.1\% of the snapshots for the diffusion advection
equation.  Whenever we refer to the reorder approach, we assume that we
use a reorder basis with $3$ additional snapshots.

\subsection{Runtime}

\noindent
We compare the runtime of simulation runs for the different approaches
on different mobile devices for both, single queries and multiple
queries.  Runtime of the adaptive approach can be split into runtime for
the local case, when the available reduced basis provides sufficient
quality, and runtime for the remote case, when the reduced basis needs
an update from the server.  The subspace approach was evaluated using
$83.5\,\%$ of the snapshots.  For each skipped snapshot, it had to
compute the residual and the subspace.  For the reorder approach, we
assume that a reorder basis was generated and that the approach only
needs 69.1\,\% of the snapshots.  We included all computations for the
overhead.  For the mobile-approach, we used two implementations, one
using pure Java and one using the Android Native Development Kit (NDK)
in C++.  We repeated the measurements for different discretizations of
the underlying full simulation problem and for different numbers of
snapshots.

\begin{figure}
  \centering
  \includegraphics[width=0.8\linewidth]{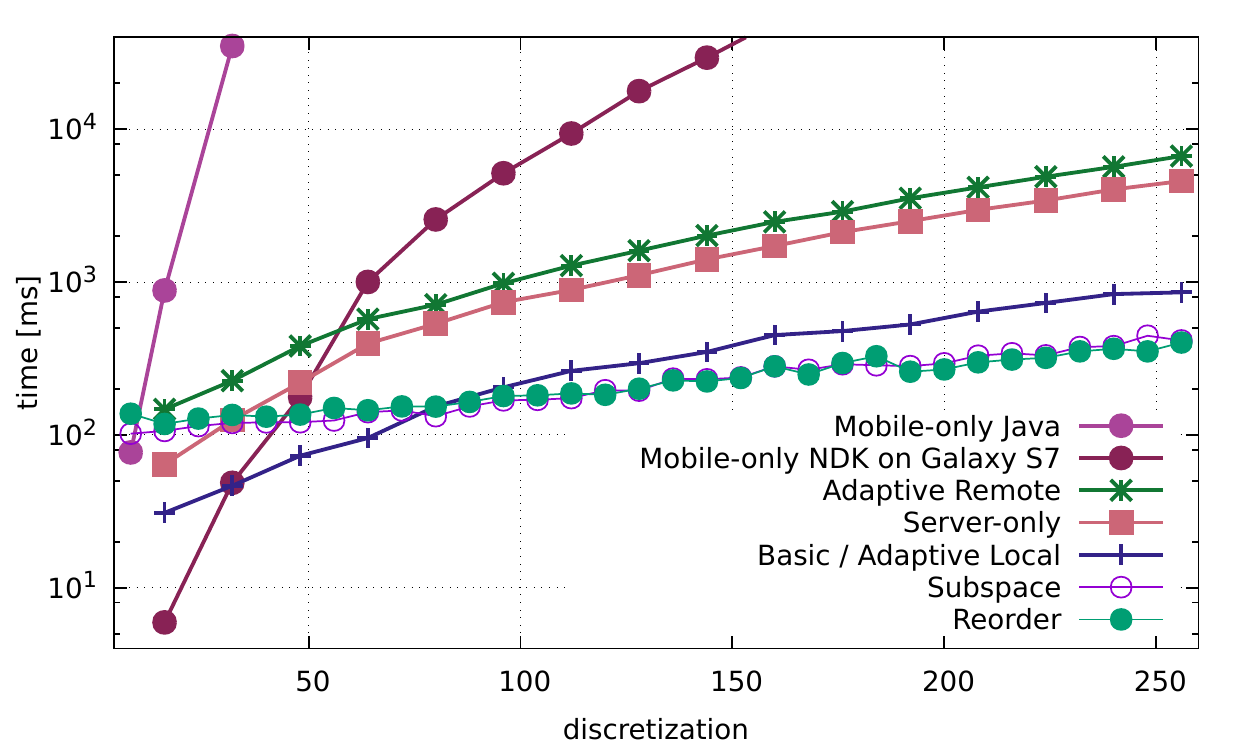}
  \caption{%
    Runtime for varying problem discretizations.
  }
  \label{fig:runtime-dimension}
\end{figure}

Figure~\ref{fig:runtime-dimension} depicts the average runtime for the
processing of single queries for different sizes of the full problem.
The full problem discretization is equidistant on both axes of the 2D
domain.  Therefore, for instance, for discretization $\mathcal{D} = 32$,
a matrix equation with a $32^2 \times 32^2$ matrix has to be solved.
The used reduced bases had $15$ snapshots.  Most results from the Galaxy
S7 were the same as for the Note 4.  Only the performance of the
mobile-only approach using the NDK was significantly better on the
Galaxy S7.  For simplicity, only the better results from the S7 are
depicted.  All other results depicted in
Fig.~\ref{fig:runtime-dimension} are from the Note 4.  Results from the
local case of the adaptive approach are very similar to results of the
basic approach.  Therefore, only the remote case of the adaptive
approach is depicted.  The server-only approach is over 280 times faster
than the mobile-only approach in pure Java.  The basic approach is again
over 5 times faster than the server-only approach.  The subspace
approach is over $51\,\%$ faster than the basic approach.  The reorder
approach is only 3\,\% faster than the subspace approach.  This is
partly caused by the random access on the data file, which can be read
as one big bulk operation in the subspace approach.

\begin{figure}[t]
  \centering
  \begin{subfigure}{.49\linewidth}
    \includegraphics[width=\columnwidth]{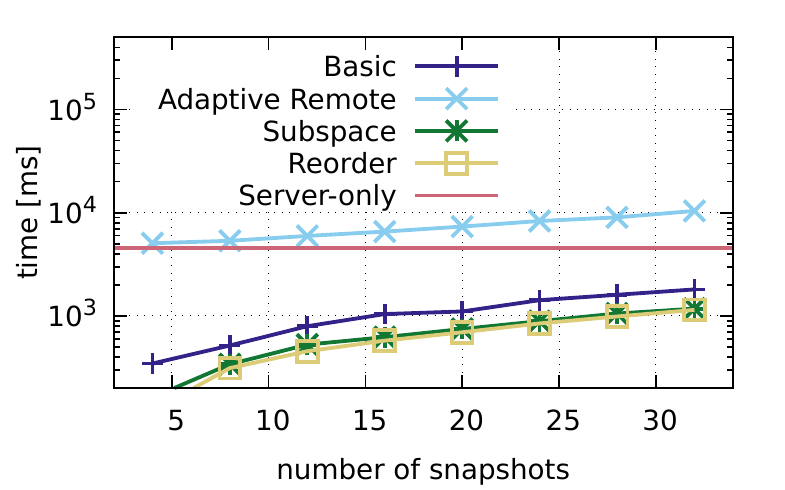}
    \caption{%
      Runtime for single queries with varying snapshot number.
    }
    \label{fig:runtime-snapshots}
  \end{subfigure}\hfill
  \begin{subfigure}{.49\linewidth}
    \includegraphics[width=\columnwidth]{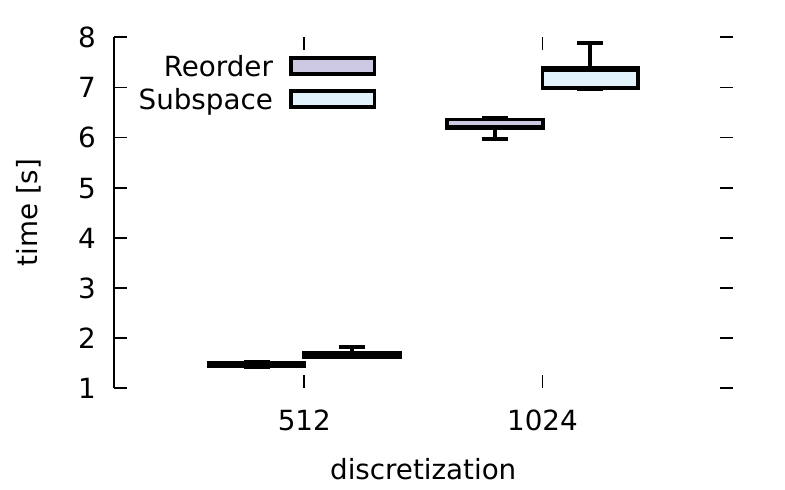}
    \caption{%
      Runtime for very high dimensions of 512 and 1024 with 15
      snapshots.
    }
    \label{fig:runtime-dimension-big}
  \end{subfigure}\hfill
  \caption{Evaluation of the runtime.}
\end{figure}

As the improvement of the reorder approach until dimension $\mathcal D =
256$ is not significant, we evaluated the subspace and the reorder
approach for bigger reduced problems of dimension $\mathcal D = 512$ and
$\mathcal D = 1024$.  While for the previous evaluation, snapshots had
sizes of up to 500\,KB, for dimension $\mathcal D = 1024$, one snapshot
has 8\,MB.  Figure~\ref{fig:runtime-dimension-big} depicts results of
this evaluations, where the reorder approach is 10\,\% faster for
dimension $\mathcal D = 512$ and 16.5\,\% faster for dimension 1024
comparing the median execution times.

As both implementations of the mobile-only approach perform very poorly,
we compare our approaches in the following only with the server-only
approach.

\begin{figure}
  \begin{subfigure}{.49\linewidth}
    \includegraphics[width=\columnwidth]{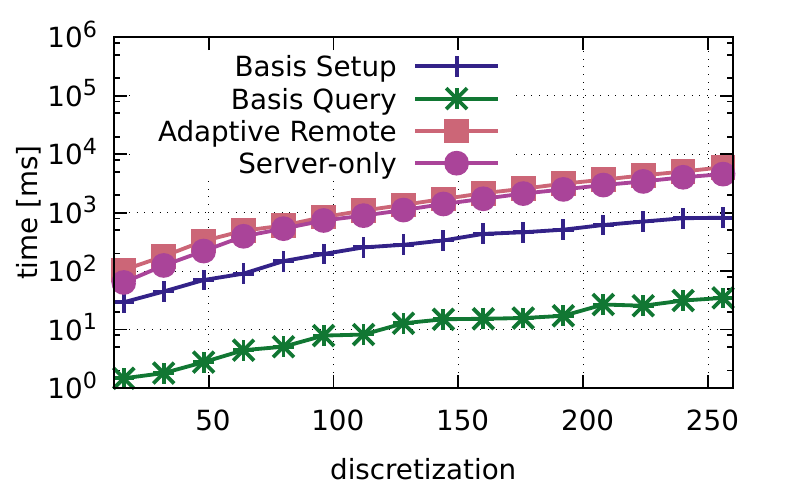}
    \caption{%
      Runtime for multiple queries with varying full problem dimension.
    }
    \label{fig:multi-runtime-dimension}
  \end{subfigure}\hfill
  \begin{subfigure}{.49\linewidth}
    \includegraphics[width=\columnwidth]{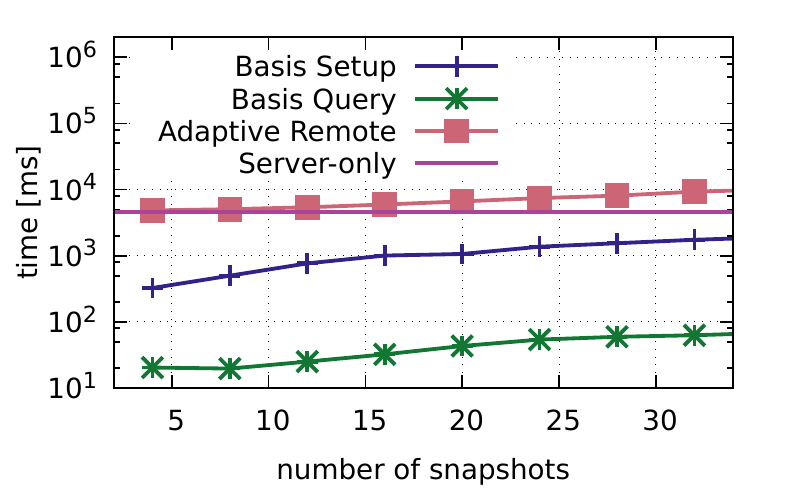}
    \caption{%
      Runtime for multiple queries with varying snapshots number.
    }
    \label{fig:multi-runtime-snapshots}
  \end{subfigure}
\end{figure}

Next, we compare the runtime for processing single queries with varying
number of snapshots in the reduced basis.
Figure~\ref{fig:runtime-snapshots} depicts the results with full problem
dimension $\mathcal{D} = 256$.  As the server-only approach computes the
full problem, it does not depend on the snapshot size.  With growing
number of snapshots, our approaches need more time.  The speedup of the
basic approach against the server-only approach is over $13.2$ for $4$
snapshots and decreases to $2.5$ for $32$ snapshots. However, with $64$
snapshots, the speedup of our approaches against the server-only
approach is still above $1.3$ for the basic approach.  The subspace
approach is 45\,\% faster than the basic approach.  As the reorder
approach is only 2.8\,\% faster than the subspace approach, we see that
the number of snapshots does not have too much impact on the performance
of the reorder approach when snapshots are small.

Many applications need multiple queries, e.g., to continuously visualize
simulation results for augmented reality.  Our basic and adaptive
approach can be split into a setup part, where the snapshots are loaded
from internal storage, and a query part, where the solution for a
parameter $\mu$ is computed.  Figure~\ref{fig:multi-runtime-dimension}
depicts the runtime for setup phase and processing of queries for
different sizes of the full problem with $15$ snapshots and
Fig.~\ref{fig:multi-runtime-snapshots} for different number of snapshots
with discretization $\mathcal{D} = 256$.  Both figures show that most
time is needed for the setup phase.  Processing queries even with $32$
snapshots and $\mathcal{D} = 256$ only takes 63\,ms.  Query processing
using the basic approach on the mobile device is over $131$ times faster
than the server-only approach for $\mathcal{D} = 256$ and $15$
snapshots.

\subsection{Energy Consumption}

\noindent
Energy is a very important resource for battery-powered mobile devices.
Therefore, we evaluated the energy consumption of all four approaches
with varying full-problem discretization, as well as for varying number
of snapshots.

\begin{figure}[t]
  \centering
  \begin{subfigure}{.49\linewidth}
    \includegraphics[width=\linewidth]{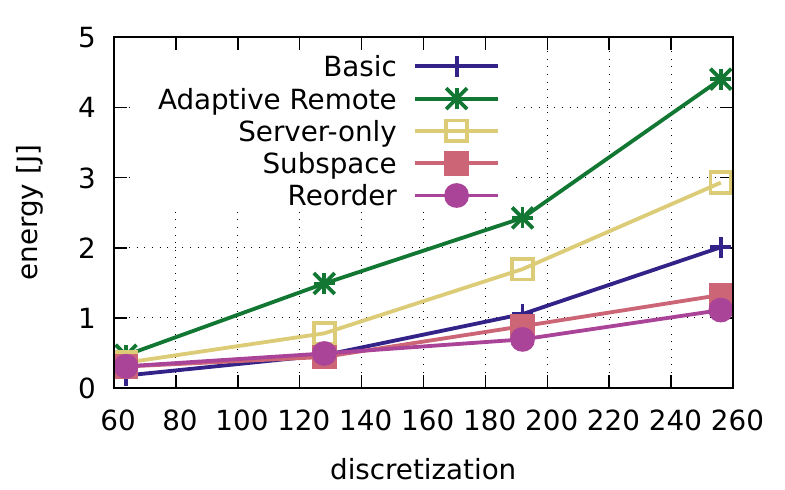}
    \caption{
      Energy consumption for different discretizations of the full
      problem.
    }
    \label{fig:energy-dimension}
  \end{subfigure}\hfill
  \begin{subfigure}{.49\linewidth}
    \includegraphics[width=\linewidth]{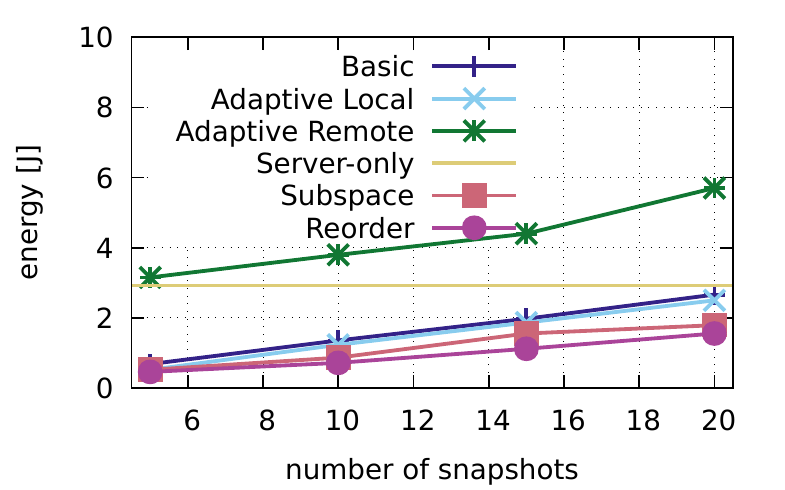}
    \caption{%
      Energy consumption for different number of snapshots.
    }
    \label{fig:energy-snapshots}
  \end{subfigure}
  \caption{Evaluation of the energy consumption.}
\end{figure}

\begin{figure}
  \centering
  \includegraphics[width=0.49\linewidth]{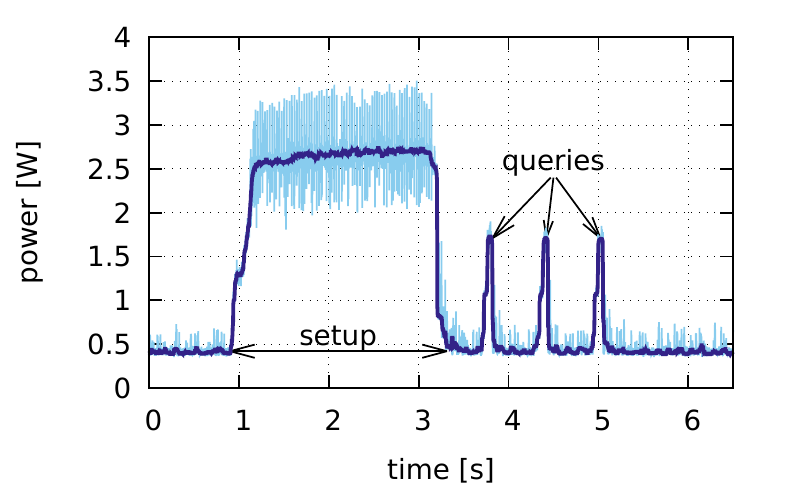}
  \caption{
    Energy consumption of the basic approach processing three queries
    after initial setup.
  }
  \label{fig:multi-solve-energy}
\end{figure}

Figure~\ref{fig:energy-dimension} depicts the energy consumption for
varying discretizations of the full problem.  The initial bases had $15$
snapshots.  Updates in the adaptive approach consume more energy than
the server-only approach.  The energy consumption of the local case of
the adaptive approach is very similar to the basic approach.  Both only
need 68\,\% of energy compared to the server-only approach.  The
subspace approach needs 34\,\% of the energy of the basic approach,
while the reorder approach saves another 18\,\% of energy compared to
the subspace approach.  The mobile-only approaches, which are not
depicted, consume significantly more energy.  The NDK version consumed
3.9\,J for discretization $\mathcal{D} = 32$ and over 84\,J for
$\mathcal{D} = 64$.  The pure Java implementation consumes more than
80\,J for $\mathcal{D} = 32$.

In addition to varying the discretization of the underlying full
problem, we also evaluated the impact of the basis size on the energy
consumption of our approaches.  Figure~\ref{fig:energy-snapshots}
depicts the energy consumption for different numbers of snapshots with
full problem size $D = 256$.  As already seen for the runtime, also the
energy consumption increases for higher number of snapshots.  The
server-only and the mobile-only approaches are not affected by the
number of snapshots.

Our approaches can reduce the energy consumption for single queries
significantly, especially when the discretization of the full problem is
high and the number of snapshots needed is low.  The adaptive approach
consumes lesser energy than the server-only approach, if more than
$8\,\%$ of the queries can be answered locally for $\mathcal{D} = 256$
and $5$ snapshots.  If the snapshot size is increased to $15$, the
adaptive approach is still beneficial if more than $58\,\%$ of  the
requests can be answered locally.  The subspace approach saves over
$32\,\%$ of energy compared to the basic approach with $20$ snapshots.
In addition, the reorder approach saves another 13\,\% compared to the
subspace approach.

We also considered the energy consumption for multiple queries for the
basic and the adaptive approach.  Figure~\ref{fig:multi-solve-energy}
depicts the power during the setup phase and the queries for a basis
with $64$ snapshots.  Between the operations, the device was idle.  Most
energy is needed for reading the reduced basis from internal storage.
After the basis is available in memory, processing one query only takes
less than 0.17\,J.  For a basis with 15 snapshots and $\mathcal{D} =
256$, the median energy consumption for one query was 0.04\,J.  This is
73 times less energy as for the server-only approach.

Overall, the evaluation showed that our approaches significantly improve
latency and energy consumption especially for processing queries after
the reduced basis is available in memory.  In this case, the basic
approach achieves a speedup of over 131 compared to the server-only
approach.  At the same time, it saves 73 times of the energy.  For
single queries, the subspace approach reduces the energy consumption in
the setup phase and therefore needs 34\,\% lesser energy as the basic
approach.  Additionally, our reorder approach is able to save again 18\%
of energy compared to the subspace approach and over 62\,\% of energy
compared to the server-only approach.


\section{Related Work}
\label{sec:relatedwork}

\noindent
In this section, we discuss work related to our approach.  Related work
can be categorized into approaches using the Reduced Basis Method (RBM),
quality-aware approaches, and code-offloading approaches.

We are not the first to execute the RBM on constrained computing
devices.  Huynh et al. already proposed to use the RBM for deployment of
thin computing platforms~\cite{Huynh2010}.  In this approach, the
pre-computation of the reduced basis is executed prior to deployment, and
the approximation using the reduced basis is executed after deployment.
However, in contrast to our approaches, they do not consider the
networking capabilities of the devices.  Therefore, their approach is
restricted to one single reduced basis that cannot be changed after the
deployment. Especially when the parameter region of queries of the
application changes over time, their approach is not able to provide
approximate solutions with quality constraints, in contrast to our
approach, 

Recently, Pandey et al. proposed a mobile distributed framework for
quality-aware applications~\cite{Pandey2016}.  Their approach adapts the
quality of computations in order to achieve better resource efficiency
of pervasive mobile applications.  To this end, they construct workflows
that reduce the quality of the application and meet the requirements of
the user.  However, we argue that numerical simulations need a more
specific method such as the reduced basis method together with specific
quality metrics like the residual.

The basic idea of approximate computing is to reduce the accuracy of
calculations in favor of reduced energy consumption, runtime, less
powerful hardware, etc. For instance, Xu et al. presented an approach
in~\cite{Xu2016} that reduces the refresh rate of memory to save energy,
which at the same time increases the probability of bit errors. A second
example is the IMPACT system by Gupta et al.~\cite{Gupta2011} that
implements an imprecise adder for low-power approximate computing.  Such
hardware-centric solutions are complementary to our approach.

Code offloading is a generic method to offload compute-intensive code
from the mobile device to a server infrastructure. Similar to our
approach, code offloading tries to optimize energy efficiency and
execution time of mobile applications. To this end, various offloading
approaches have been proposed in the literature~\cite{Cuervo2010,
Kosta2012, Yang2013, Berg2014, Berg2015, Zhang2015, Berg2016}, which try to find
an optimal partitioning of the application code into code executed
locally on the mobile device or remotely. For instance, Cuervo et al.
presented the MAUI system, which performs code offloading on a
function-level in order to optimize for energy~\cite{Cuervo2010}.  In
this approach, the user has to annotate the code in order to mark
functions that can be offloaded to the remote server.  During compile
time, the system creates proxies, which represent either local or remote
execution.  During runtime, the system continuously monitors the program
and network characteristics and decides if the function should be called
locally or remotely, depending on the energy characteristics of the
device.

Since offloading is agnostic to the semantics of the application whose
code is offloaded, it is not restricted to mobile simulations but can be
used to optimize compute-intensive mobile applications in general.
However, as a generic approach code offloading is not optimized for the
specific properties of numeric simulations. In particular, it does not
consider the quality of the simulation result and does not exploit the
possibility to trade-off quality for energy efficiency as we do by using
the Reduced Basis Method.


\section{Conclusion \& Future Work}
\label{sec:conclusion}

\noindent
In this paper, we presented a middleware for enabling complex numerical
simulations on resource-constrained mobile devices by distributing the
simulation between mobile device and a server infrastructure.  Such a
middleware is needed for interactive simulations in the field, e.g., an
engineer using a head-mounted augmented reality device who wants to
simulate the heat in an object to adjust its placement according to the
surrounding materials.  We presented four approaches for solving this
problem using the Reduced Basis Method (RBM), which pre-computes a
reduced representation of the simulation to reduce the evaluation time.
The first approach was to pre-compute a reduced basis on the server and
send this basis to the mobile device.  In order to calculate an
approximation of the simulation, no further communication is necessary.
The second approach was more interactive and utilized the fast error
indicator of the RBM.  Using this indicator, the mobile device can
efficiently check if the quality demands of the application are
fulfilled.  If the quality is not sufficient, the mobile device requests
a basis update from the server.  After such an update, the mobile device
is able to answer queries with similar parameters completely autonomous
without communication with the server.  Goal of the third and fourth
approach is to reduce the number of data to be read from internal
storage, which we identified as major energy consumer.  In addition, we
also presented a novel approach for the pre-computation, which further
reduces the data needed from internal storage during runtime.

We evaluated our approaches on real mobile devices in a real wireless
network.  We showed that our approach has lower energy consumption and
is multiple times faster compared to two simple approaches.  In
particular, we showed that our approach, once it has performed a setup
phase, is over 131 times faster and consumes 73 times less energy
compared to offloading everything to a connected server.  Still, our
approach keeps quality requirements as requested and reports an error
indicator to the user.

In the future, we will further extend our approach by integrating real
time sensor data streams into the simulation.


\section*{Acknowledgments}

\noindent
The authors would like to thank Felix Baumann for his help with the
energy measurement equipment and Dominik Schreiber for his help in the
evaluation.  Furthermore, the authors would like to thank the German
Research Foundation (DFG) for financial support of the project within
the Cluster of Excellence in Simulation Technology (EXC 310/2) at the
University of Stuttgart.

\bibliographystyle{elsarticle-num}
\bibliography{literature}

\begin{thebibliography}{10}
\expandafter\ifx\csname url\endcsname\relax
  \def\url#1{\texttt{#1}}\fi
\expandafter\ifx\csname urlprefix\endcsname\relax\def\urlprefix{URL }\fi
\expandafter\ifx\csname href\endcsname\relax
  \def\href#1#2{#2} \def\path#1{#1}\fi

\bibitem{Dibak2014}
C.~Dibak, B.~Koldehofe, {Towards Quality-aware Simulations on Mobile Devices},
  in: Informatik 2014, Gesellschaft f{\"u}r Informatik (GI), 2014.

\bibitem{Dibak2015}
C.~Dibak, F.~D{\"u}rr, K.~Rothermel, {Numerical Analysis of Complex Physical
  Systems on Networked Mobile Devices}, in: MASS, IEEE, 2015.
\newblock \href {http://dx.doi.org/10.1109/MASS.2015.12}
  {\path{doi:10.1109/MASS.2015.12}}.

\bibitem{Dibak2017b}
C.~Dibak, F.~D{\"u}rr, K.~Rothermel, Demo: Server-assisted interactive mobile
  simulations for pervasive applications, in: PerCom Workshops, 2017.
\newblock \href {http://dx.doi.org/10.1109/PERCOMW.2017.7917525}
  {\path{doi:10.1109/PERCOMW.2017.7917525}}.

\bibitem{Ware2016}
J.~Ware, N.~Roy, An analysis of wind field estimation and exploitation for
  quadrotor flight in the urban canopy layer, in: ICRA, IEEE, 2016, pp.
  1507--1514.
\newblock \href {http://dx.doi.org/10.1109/ICRA.2016.7487287}
  {\path{doi:10.1109/ICRA.2016.7487287}}.

\bibitem{Huynh2010}
D.~Huynh, D.~Knezevic, J.~Peterson, A.~Patera, High-fidelity real-time
  simulation on deployed platforms, Computers \& Fluids 43, symposium on High
  Accuracy Flow Simulations.
\newblock \href {http://dx.doi.org/10.1016/j.compfluid.2010.07.007}
  {\path{doi:10.1016/j.compfluid.2010.07.007}}.

\bibitem{Dibak2017a}
C.~Dibak, A.~Schmidt, F.~D{\"u}rr, B.~Haasdonk, K.~Rothermel, Server-assisted
  interactive mobile simulations for pervasive applications, in: PerCom, IEEE,
  2017.
\newblock \href {http://dx.doi.org/10.1109/PERCOM.2017.7917857}
  {\path{doi:10.1109/PERCOM.2017.7917857}}.

\bibitem{Patera2007}
A.~T. Patera, G.~Rozza, Reduced basis approximation and a posteriori error
  estimation for parametrized partial differential equations, version 1.0,
  Copyright MIT 2006, to appear in (tentative rubric) MIT Pappalardo Graduate
  Monographs in Mechanical Engineering (2007).

\bibitem{Haasdonk2011b}
B.~Haasdonk, M.~Ohlberger, Efficient reduced models and a posteriori error
  estimation for parametrized dynamical systems by offline/online
  decomposition, Mathematical and Computer Modelling of Dynamical Systems
  17~(2) (2011) 145--161.
\newblock \href {http://dx.doi.org/10.1080/13873954.2010.514703}
  {\path{doi:10.1080/13873954.2010.514703}}.

\bibitem{Haasdonk2014}
B.~Haasdonk, Reduced basis methods for parametrized {PDEs} -- a tutorial
  introduction for stationary and instationary problems, {C}hapter in P.
  Benner, A. Cohen, M. Ohlberger and K. Willcox: "Model Reduction and
  Approximation for Complex Systems", SIAM, Philadelphia (2016).

\bibitem{Barrault2004}
M.~Barrault, Y.~Maday, N.~C. Nguyen, A.~T. Patera, An 'empirical interpolation'
  method: application to efficient reduced-basis discretization of partial
  differential equations, Comptes Rendus Mathematique 339~(9) (2004) 667--672.
\newblock \href {http://dx.doi.org/10.1016/j.crma.2004.08.006}
  {\path{doi:10.1016/j.crma.2004.08.006}}.

\bibitem{Veroy2003}
K.~Veroy, C.~Prud'Homme, D.~V. Rovas, A.~T. Patera, A posteriori error bounds
  for reduced-basis approximation of parametrized noncoercive and nonlinear
  elliptic partial differential equations, in: Proceedings of the 16th AIAA
  computational fluid dynamics conference, Vol. 3847, Orlando, Florida, 2003,
  pp. 23--26.
\newblock \href {http://dx.doi.org/10.2514/6.2003-3847}
  {\path{doi:10.2514/6.2003-3847}}.

\bibitem{Rozza2008}
G.~Rozza, D.~B.~P. Huynh, A.~T. Patera,
  \href{https://doi.org/10.1007/s11831-008-9019-9}{Reduced basis approximation
  and a posteriori error estimation for affinely parametrized elliptic coercive
  partial differential equations}, Archives of Computational Methods in
  Engineering 15~(3) (2008) 229.
\newblock \href {http://dx.doi.org/10.1007/s11831-008-9019-9}
  {\path{doi:10.1007/s11831-008-9019-9}}.
\newline\urlprefix\url{https://doi.org/10.1007/s11831-008-9019-9}

\bibitem{Pandey2016}
P.~Pandey, D.~Pompili, Mobidic: Exploiting the untapped potential of mobile
  distributed computing via approximation, in: 2016 IEEE International
  Conference on Pervasive Computing and Communications (PerCom), 2016, pp.
  1--9.
\newblock \href {http://dx.doi.org/10.1109/PERCOM.2016.7456515}
  {\path{doi:10.1109/PERCOM.2016.7456515}}.

\bibitem{Xu2016}
Q.~Xu, T.~Mytkowicz, N.~Kim, Approximate computing: A survey, Design \& Test,
  IEEE 33~(1) (2016) 8--22.
\newblock \href {http://dx.doi.org/10.1109/MDAT.2015.2505723}
  {\path{doi:10.1109/MDAT.2015.2505723}}.

\bibitem{Gupta2011}
V.~Gupta, D.~Mohapatra, S.~P. Park, A.~Raghunathan, K.~Roy, Impact: Imprecise
  adders for low-power approximate computing, in: ISLPED, IEEE, 2011.

\bibitem{Cuervo2010}
E.~Cuervo, A.~Balasubramanian, D.-k. Cho, A.~Wolman, S.~Saroiu, R.~Chandra,
  P.~Bahl, Maui: making smartphones last longer with code offload, in: MobiSys,
  ACM, 2010.
\newblock \href {http://dx.doi.org/10.1145/1814433.1814441}
  {\path{doi:10.1145/1814433.1814441}}.

\bibitem{Kosta2012}
S.~Kosta, A.~Aucinas, P.~Hui, R.~Mortier, X.~Zhang, Thinkair: Dynamic resource
  allocation and parallel execution in the cloud for mobile code offloading,
  in: INFOCOM, IEEE, IEEE, 2012.

\bibitem{Yang2013}
S.~Yang, Y.~Kwon, Y.~Cho, H.~Yi, D.~Kwon, J.~Youn, Y.~Paek, Fast dynamic
  execution offloading for efficient mobile cloud computing, in: PerCom, 2013.
\newblock \href {http://dx.doi.org/10.1109/PerCom.2013.6526710}
  {\path{doi:10.1109/PerCom.2013.6526710}}.

\bibitem{Berg2014}
F.~Berg, F.~D{\"u}rr, K.~Rothermel, Optimal predictive code offloading, in:
  MobiQuitous, ICST, 2014.
\newblock \href {http://dx.doi.org/10.4108/icst.mobiquitous.2014.258023}
  {\path{doi:10.4108/icst.mobiquitous.2014.258023}}.

\bibitem{Berg2015}
F.~Berg, F.~D\"urr, K.~Rothermel, in: WiMob, IEEE, 2015.
\newblock \href {http://dx.doi.org/10.1109/WiMOB.2015.7348013}
  {\path{doi:10.1109/WiMOB.2015.7348013}}.

\bibitem{Zhang2015}
Y.~Zhang, D.~Niyato, P.~Wang, Offloading in mobile cloudlet systems with
  intermittent connectivity, Transactions on Mobile Computing\href
  {http://dx.doi.org/10.1109/TMC.2015.2405539}
  {\path{doi:10.1109/TMC.2015.2405539}}.

\bibitem{Berg2016}
F.~Berg, F.~D{\"u}rr, K.~Rothermel, {Increasing the Efficiency of Code
  Offloading in n-tier Environments with Code Bubbling}, in: MobiQuitous, IEEE,
  2016.
\newblock \href {http://dx.doi.org/10.1145/2994374.2994375}
  {\path{doi:10.1145/2994374.2994375}}.

\end{thebibliography}

\appendix

\section{Error Estimation for the Reduced Basis Method}
\label{sec:error-estimation}

\noindent
In simulations, the trade-off between accuracy and computational effort
is made on many levels.  One key property for serious simulation
applications is to estimate or indicate the error.  One important tool
for error indication is the residual.  The residual is the difference in
an equation to be solved.  For the RBM, the residual can be computed
very fast, as essential parts can be pre-computed.  As this is used in
the adaptive and the subspace approach, we shortly sketch the procedure.

The RBM provides an approximation $V u_V(\mu)$ for the solution $u(\mu)$
in the equation ${A(\mu) u(\mu) = f(\mu)}$ (c.f. Section~\ref{sec:rbm}).
We can write this equation as $A(\mu) V u_V(\mu) = f(\mu) + r$ with
residual $r$.  Using this equation, we can compute ${\lVert r\rVert^2 =
r^T r}$ as follows
\begin{subequations}
  \label{align:fast-residual}
  \begin{align}
    \lVert \mathbf{r}\rVert^2 =\ & \mathbf{u}_V(\mu)^T \underbrace{%
      \mathbf{V}^T \mathbf{A}(\mu)^T \mathbf{A}(\mu) \mathbf{V}
    }_{\mathbf{r}_1} \mathbf{u}_V(\mu) \label{align:fast-residual-a}\\
    & - \mathbf{u}_V(\mu)^T \underbrace{\mathbf{V}^T \mathbf{A}(\mu)^T
    \mathbf{f}(\mu)}_{\mathbf{r}_2} \\
    & - \underbrace{\mathbf{f}(\mu)^T \mathbf{A}(\mu)
    \mathbf{V}}_{\mathbf{r}_3} \mathbf{u}_V(\mu) \\
    & + \underbrace{\mathbf{f}(\mu)^T \mathbf{f}(\mu)}_{r_4}.
  \end{align}
\end{subequations}
Notice that $r_1$ to $r_4$ can be expressed as separable matrices. These
matrices can be pre-computed after the basis construction.  Also notice
that a new snapshot will increase the matrix $V$ by exactly one column
vector.  Therefore, this will only add columns and rows to the matrices
$r_1$ to $r_4$.  Analogously, the last snapshot can be removed from the
residual computation by trimming the matrices as needed for the subspace
approach.



\end{document}